\documentclass[sigconf, nonacm]{acmart}
\usepackage{epigraph}
\usepackage{mathrsfs}
\usepackage{enumitem}
\usepackage{makecell} 
\usepackage{subfig}
\usepackage{multirow}

\usepackage{mathtools}   
\usepackage{makebox}
\usepackage{caption}
\usepackage{booktabs}
\usepackage{colortbl}
\usepackage{amsfonts} % for math fonts
\usepackage{url}
\usepackage{textcomp}
\usepackage{graphicx}
\usepackage[breakable, theorems, skins]{tcolorbox}
\usepackage{float}
\restylefloat{table}

\newcommand\vldbdoi{10.1145/3807515}

\newcommand\vldbvolume{69}
\newcommand\vldbissue{6}
\newcommand\vldbyear{2026}
\newcommand\vldbauthors{\authors}
\newcommand\vldbtitle{\shorttitle} 
\newcommand\vldbavailabilityurl{}
\newcommand\vldbpagestyle{plain} 

\newcommand{\gdpr}{$\mathscr{G}$}
\newcommand{\sref}[1]{Section-\ref{#1}}
\newcommand{\vheading}[1]{\vspace{0.05in}\noindent\textbf{#1}}

\begin{document}
\title{Faults and Pitfalls in Implementing the Right to be Forgotten}

%%
%% The "author" command and its associated commands are used to define the authors and their affiliations.

\author{Chen Sun}
\affiliation{%
  \institution{\it \normalsize Computer Science}
  \institution{University of Iowa}
}

\author{Nikolas Guggenberger}
\affiliation{%
\institution{\it \normalsize Law School}
  \institution{University of Houston}
%  \institution{ }
}

\author{Supreeth Shastri}
\affiliation{%
  \institution{\it \normalsize Computer Science}
  \institution{University of North Texas}
}
\email{}

\begin{abstract}
Right to be Forgotten (RTBF) in one of the oldest and prominent of the legal data rights. While its legal intention is straight forward (for example, the GDPR describes it in just 417 words), the computing community has found it challenging to implement this in practice. For example, regulators have issued 205 RTBF violations in the first five years of GDPR i.e., an RTBF failure once every 9 days, on average. In this work, we identify the uncertainties and risks in supporting RTBF from a computing perspective. Then, to mitigate these challenges, we propose a two-phase approach that bridges an intrinsic dichotomy between law and computing. We demonstrate the effectiveness of our technique by showing how it could have fully avoided 80\% of RTBF violations that occurred in the year-6 of GDPR. We also discover six long-standing practices of computing and data management that have become anti-patterns for RTBF. Finally, to ground our research, we introduce RTBF capability into Elasticsearch, a popular open-source search engine. 
\end{abstract}

\maketitle

\pagestyle{\vldbpagestyle}
\begingroup\small\noindent\raggedright\textbf{Reference Format:}\\
\vldbauthors. \vldbtitle. Communications of the ACM, \vldbvolume(\vldbissue): \vldbyear.\\
\href{https://doi.org/\vldbdoi}{doi:\vldbdoi}
\endgroup
\begingroup
\renewcommand\thefootnote{}\footnote{\noindent
This work is licensed under the Creative Commons BY-NC-ND 4.0 International License. Visit \url{https://creativecommons.org/licenses/by-nc-nd/4.0/} to view a copy of this license. This is an extended version of the article that appeared in CACM with the aforementioned DOI. Copyright is held by the owner/author(s). Publication rights licensed to the Communications of the ACM. \\
\href{https://doi.org/\vldbdoi}{doi:\vldbdoi} \\
}\addtocounter{footnote}{-1}\endgroup

\ifdefempty{\vldbavailabilityurl}{}{
\vspace{.3cm}
\begingroup\small\noindent\raggedright\textbf{PVLDB Artifact Availability:}\\
The source code, data, and/or other artifacts have been made available at \url{\vldbavailabilityurl}.
\endgroup
}

\section{Introduction}
\label{sec:intro}

\setlength{\epigraphwidth}{2.4in}
\setlength{\epigraphrule}{0.1pt}
\epigraph{\emph{``For every complex problem, there is an answer that is clear, simple, and wrong.''}}{ Henry Louis Mencken}

%\vspace{1mm}
Ancient wisdom says that everything that has a beginning has an ending. However, when it comes to the lifecycle of personal data, the ending was nowhere in sight. In fact, for much of its existence, the computing community has evolved without treating deletion as a first-class operation. This practice had to change when the European Union introduced the Right to be Forgotten (RTBF)---first in 2014, as a stand-alone right applicable only to online search engines and then in 2018, as a universal right applicable to all data controllers through the General Data Protection Regulation (GDPR)~\cite{gdpr-regulation}. 

\vspace{1mm}
%\emph{``Isn't it just deletion?''} has been the computing community's standard reaction to the requirements of RTBF. While the end goal of RTBF is indeed the deletion of data, casting RTBF as \emph{just deletion} is akin to saying that eating is only for nutrition. It is not surprising that over the last five years, an RTBF penalty is issued once every eight days---a clear sign that the computing and data management communities have continued to oversimplify and misunderstand the nuances of RTBF. Our work is an attempt to remedy this disconnect.

\emph{``Isn't it just deletion?''} has been the computing community's standard reaction to the requirements of RTBF. While the end goal of RTBF is indeed the deletion of data, casting RTBF as \emph{just deletion} is akin to saying that eating is only for nutrition or sex is just about reproduction. It is not surprising that over the first five years of GDPR, an RTBF penalty is issued once every nine days---a clear sign that the computing and data management communities have continued to oversimplify, misunderstand, and ill-implement RTBF. Our work is an attempt to remedy this disconnect.

\vspace{1mm}
We demonstrate how RTBF exposes computing systems to uncertainties and challenges at all stages of design and operation (in \sref{sec:rtbf-challenges}), and how RTBF has invalidated principles and practices of data management with decades of precedent (in \sref{sec:RTBF-anti-patterns}). To mitigate these impact, we propose a principled approach for introducing RTBF capability in computing systems (in \sref{sec:design}). Our solution is rooted in two key insights: (i) the need to bridge an intrinsic dichotomy that exists between computing and law i.e., computing systems are created to be precise and static, but laws are written to be abstract and interprettable, and (ii) modeling compliance as a \emph{via negativa} problem i.e., instead of trying to build a perfectly compliant system, it is much easier to weed out all the known violations from our system.

\vspace{1mm}
We ground our work firmly in both law and computing---the former, by basing our analysis on GDPR and the first five years of its enforcement, and the latter, by implementing our findings in Elasticsearch, a widely deployed search engine. The interdisciplinary nature of this work has allowed us to integrate techniques from both law (for e.g., tracking the state-of-the-art and IRAC analysis) and computing (for e.g., layered design and anti-patterns), and thus be able to make evidence-based recommendations that previous domain-specific investigations have lacked. In particular, we make three key contributions:

%While RTBF is one of the seven data rights accorded by GDPR, it is considered a first amongst equals. This is because once RTBF is successfully executed, no other data rights can be exercised on the said data.

\vspace{1mm}
\begin{itemize}[leftmargin=6mm, parsep=1mm, topsep=1mm]

\item{We propose a novel two-phase approach for introducing RTBF capability. In \emph{law-driven design}, we analyze the legal requirements of RTBF and map them to specific computing and data management tasks. Then, in \emph{enforcement-driven refinement}, we analyze RTBF enforcements to identify the available design choices for those tasks and to weed out non-compliant ones.}

\item{We evaluate our approach by showing how it can avoid \emph{faults in the future} as well as how it can identify \emph{pitfalls from the past}. Specifically, our longitudinal analysis shows that our technique could have fully avoided 80\% of faults and partially avoided 20\% of faults that occurred in year-6 of GDPR. Looking back in time, we identify six \emph{RTBF anti-patterns} which are long-standing principles and practices of computing and data management systems that make it difficult to support RTBF.} 

\item{We demonstrate how to apply our technique in practice. We analyze the RTBF capabilities of Elasticsearch and identify the functionalities that are lacking. Then, we implement \emph{timed- and thorough-delete} operations in Elasticsearch, and measure its impact using the Rally benchmark. We release all software artifacts and datasets at {\color{blue}\url{https://github.com/lawfulcomputing/RTBF}}.}

\end{itemize}

\section{Background and Motivation}
\label{sec:background}

In this section, we discuss the importance of the problem, review related work, and establish the need for and novelty of our work.

\subsection{RTBF} 
\label{sec:rtbf-primer}
Right to be Forgotten, alternatively referred to as Right to Erasure, is the right of an individual person to request an organization to delete their personal data. RTBF gives people an ability to prevent others from seeing their personal information that they deem inappropriate, irrelevant, or simply withdraw consent to be used. This right is distinct from right to privacy \cite{eu-fundamental-rights}, which prevents governments and other entities from forcing a person to reveal their personal data; instead, RTBF is to be used for information that is already disclosed (either to an organization or to the public), but the individual no longer wants it to remain so.

\vspace{1mm}
RTBF became a legal right for the first time in 2014. In a case against Google \cite{ecj-rtbf-ruling}, the European Court of Justice ruled that people can request search engines to remove certain links from their search results, if their privacy concerns outweigh the public interest in the information contained in those links. Later, when GDPR went into effect in 2018, RTBF was expanded to cover all data controllers, not just search engines. Even GDPR did not make RTBF an absolute right i.e., for an RTBF request to be honored, it has to meet one of the six conditions and not fall under one of the five exemptions (we show these in Figure \ref{fig:rtbf-mapping}, where we produce GDPR article 17 verbatim). To keep our discussion concrete, we focus exclusively on GDPR's version of RTBF and prefix its articles with \gdpr.

\vspace{1mm}
The need for RTBF arose in early 21st century when organizations began collecting personal information at scale, and search engines made these accessible globally. As Viktor Mayer-Schonberger has chronicled in his book \cite{delete-book}, throughout the human history, forgetting was the norm and remembering was the exception. Given how this phenomenon gotten largely reversed in the recent decades \cite{hew-report}, RTBF is hailed a countermeasure against this trend. That said, RTBF has received criticisms \cite{rtbf-criticism-1, rtbf-criticism-2, rtbf-criticism-3, rtbf-criticism-4, rtbf-criticism-5} as a means to rewrite history, weaken the freedom of expression, enable censorship and other less desirable social outcomes. While this is an important debate for our society, the focus of our work is in exploring the challenges that computing systems face in implementing RTBF and how to systematically solve them.

\subsection{Challenges in Complying with RTBF}
\label{sec:rtbf-challenges}
When new regulations are enacted, legal experts and policy makers tend to limit their expositions to core legal principles that are broadly interpretable and will hold the test of time. While legally prudent, this strategy makes it challenging for computing systems to adapt and support regulations such as RTBF. These challenges can come in the form of lack of precise specifications, undefined tradeoffs in performance-vs-risk, uncertainties in managing new technologies, among others. We illustrate how such challenges can manifest at different stages of system design and operation: 

\vspace{1mm}
\vheading{Examples at architecture level.} Systems that are architected with little or no prior consideration for RTBF, find it hard to add that capability later. For example, an organization that uses a personal data as a primary key in their databases, would find it tricky to delete that item. Problems could also stem from unwise choices in organizing the data. For instance, Clearview AI built their facial recognition system by training on billions of images from the Internet. However, when people approached them with RTBF requests, they realized that they could not identify all the photos that belong to a given person (since they had not tagged the images at the time of collecting or processing). They were fined in 2022 by multiple regulators \cite{clearview-1, clearview-2, clearview-3} for this limitation. More generally, RTBF in machine learning systems is a nascent area of research with no generic or efficient solutions that can help models forget a select data in their training set.

\vspace{1mm}
\vheading{Examples at design and implementation level.} RTBF opens up many uncertainties and unknown tradeoffs at the systems level. Consider the \emph{latency of deletion} i.e., how soon after the request, should the data be removed. Designers could opt for a strict compliance by making deletions synchronous in real-time, or choose a relaxed compliance by allowing deletions to happen eventually. Prior work \cite{gdpr-vldb} has shown the effect of synchronous deletion on two popular database systems, Redis and PostgreSQL, both of which experienced a slowdown of up to 20\%. On the other hand, eventual compliance allows stale data to linger in the system for unspecified amount of time, posing security and privacy risks. Next, consider the \emph{depth of deletion} i.e., should the data be deleted from all memory and storage subsystems going all the way to the hardware, or simply be forgotten at the service level. While the former leads to a strict form of compliance, it adds significantly to the latency and complexity of the deletion process. The latter, however, exposes the organization to legal risks since other services may unwittingly end up using the said data. GDPR does not offer clarity on many such systems level issues.

\vspace{1mm}
\vheading{Examples at operation level.} RTBF is not an absolute right i.e., just because an RTBF request is made on valid personal data by a verified data subject, does not mean that it should be honored. GDPR requires all controllers to balance the rights of individuals with the interests and obligations to the society. This is challenging for organizations since it turns RTBF from a \emph{generic operation} to a \emph{highly individualized process}, thereby making it hard to fully automate it. The gravity of this challenge is evident when you consider that RTBF is operated largely as a manual process at Google and Microsoft \cite{google-rtbf-ccs, bing-rtbf} (organizations that are considered technologically sophisticated), and that they take about 6 days, on average, to arrive at an RTBF decision. These challenges are so pervasive that 41\% of all RTBF violations are due to incorrectly interpreting the validity of RTBF requests \cite{sun2023gdprxiv}.

\subsection{Related Work}
\vheading{RTBF in computing systems.} Several organizations including Google \cite{google-rtbf-ccs}, Microsoft \cite{bing-rtbf}, and Wikipedia \cite{wikipedia-rtbf} have shared details about how their applications support RTBF. These reports primarily focus on aggregate-level characterization of the received RTBF requests and their responses, but do not offer much details on the computing aspects of their internal RTBF implementation.

\vspace{1mm}
Orthogonal to this perspective, researchers have explored how people perceive RTBF support on social media websites \cite{habib2019empirical, minaei2022empirical}, and the challenges they face while exercising RTBF \cite{habib2020s, take2022feels, rtbf-pets2016}. This body of work focuses on human-computer interfacing for RTBF, and treats the target computing systems as black boxes. In contrast, we focus on end-to-end design and implementation of RTBF.

\vspace{1mm}
\vheading{Deletion in computing systems.} Deletion is one of the fundamental operations of database systems (as represented in \emph{CRUD}, the well-known acronym for \emph{Create-Read-Update-Delete}), yet it had long been treated as a second-class operation. However, GDPR and the onset of data rights has brought the attention of the computing community back to deletion. Google cloud publishing their deletion pipeline and guaranteeing to erase all copies of the data within 180 days of requesting \cite{google-cloud-deletion} and Facebook designing a system that can assure correctness of deletion \cite{facebook-deletion} are two practical examples. Other key advances in research include Lethe \cite{boston-deletion}, a key-value database that lets users control deletion latency; a cryptographic framework for data deletion by Garg et.al., \cite{deletion-theory}; and data systems that exhibit privacy-compliance-by-construction \cite{malte-retrofitting, malte-k9db}.

\vspace{1mm}
Given the important role training data plays in machine learning, the notion of machine unlearning i.e., making ML systems forget all they learnt from a given data, has gained traction in the AI/ML community \cite{machine-unlearning}. These work primarily focus on making the deletion efficient for certain models \cite{rtbf-federated-learning, rtbf-llm, unlearning-sgd} or specific applications \cite{kmeans-deletion, unlearn-graph}. Lastly, secure data deletion i.e., deleting data irrevocably from a physical storage, has been an active area of research in file systems, operating systems, and backup systems \cite{secure-deletion-sok, secure-deletion-pm, secure-deletion-fs, secure-deletion-os, secure-deletion-disk, secure-deletion-backup, boneh1996revocable}. 

\vspace{1mm}
A key trait of all these work is that they use RTBF as a motivation to implement deletion in a target computing system (say, databases, file systems, cloud computing, machine learning systems, etc.,), but they largely abstract out the legal and policy aspects of RTBF. A central thesis of our paper is that RTBF is not just deletion, and by ignoring the interplay between law and computing, these prior work fail to recognize important system design issues, tradeoffs and anti-patterns.

\section{Designing for RTBF}
\label{sec:design}

We propose a novel two-phase approach to design end-to-end RTBF capability. This is motivated by the dichotomy between computing and law---namely, while computing applications are created to be precise and static, laws are written to be abstract and interpretable---and our attempt to bridge this divide. In phase-1, we analyze the \emph{language of the law} (which changes rarely) and map it to a set of high-level computing tasks. In phase-2, we examine the \emph{enforcement of the law} (which evolves frequently) to identify available design choices and tradeoffs, and to weed out non-compliant ones. 

%We must note that even though our discussion is focused on one specific data right (namely, RTBF), we expect this approach to be broadly applicable to other data rights such as Right to object, Right to rectify, etc. 

%--------------------------------------------------------------------
% Identifying legal computing tasks 
%--------------------------------------------------------------------
\begin{figure*}[t]
\centering
\includegraphics[width=1\textwidth]{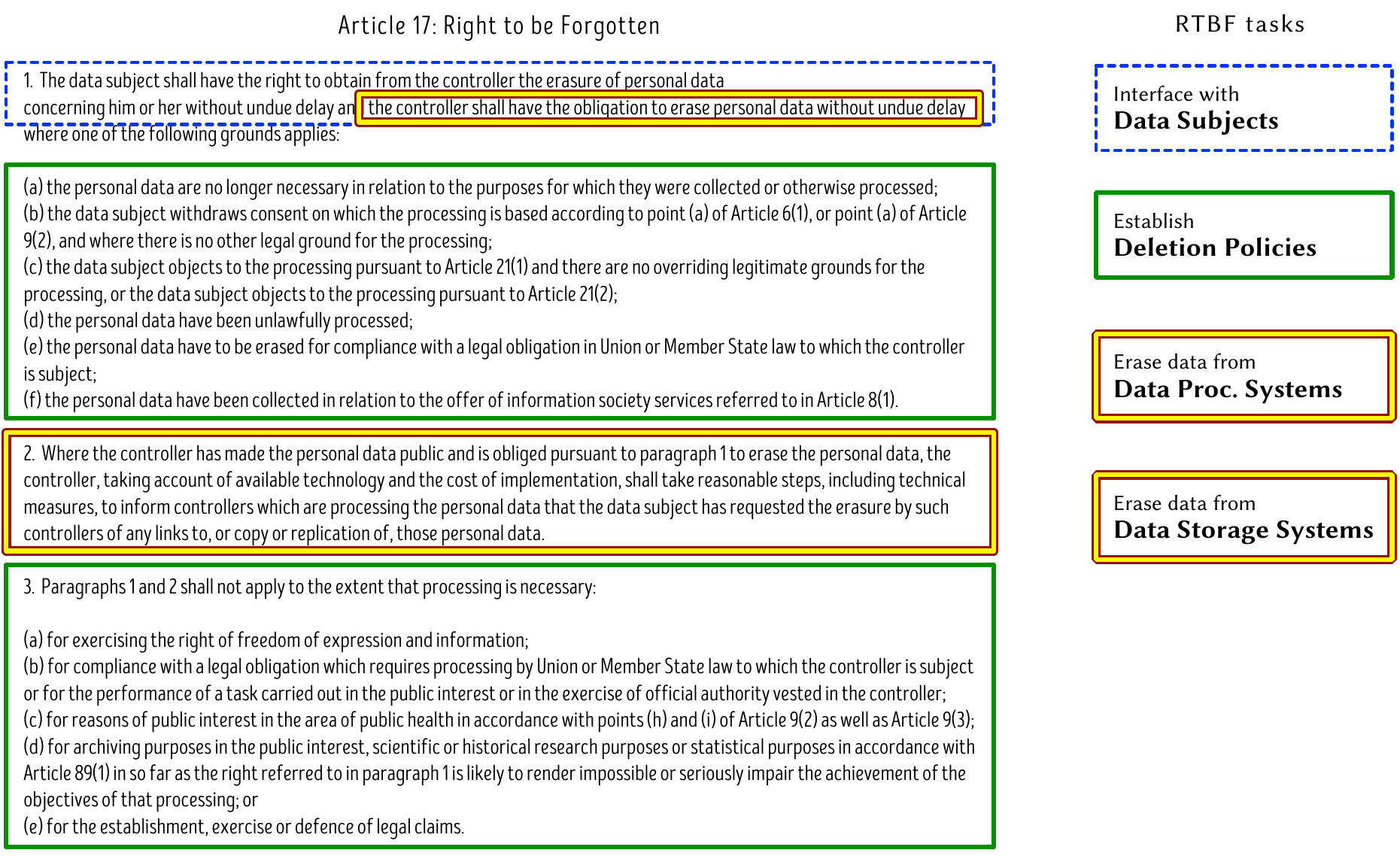}
\caption{\bf Mapping RTBF's legal requirements into computing domain. We identify four key tasks of RTBF capable systems (on the right), and the text of the regulation that requires each of these (on the left).}
\label{fig:rtbf-mapping}
\end{figure*}
%--------------------------------------------------------------------
%--------------------------------------------------------------------
% RTBF design choices
%--------------------------------------------------------------------
\begin{figure*}[]
\centering
%\vspace{5mm}
\includegraphics[width=1\textwidth]{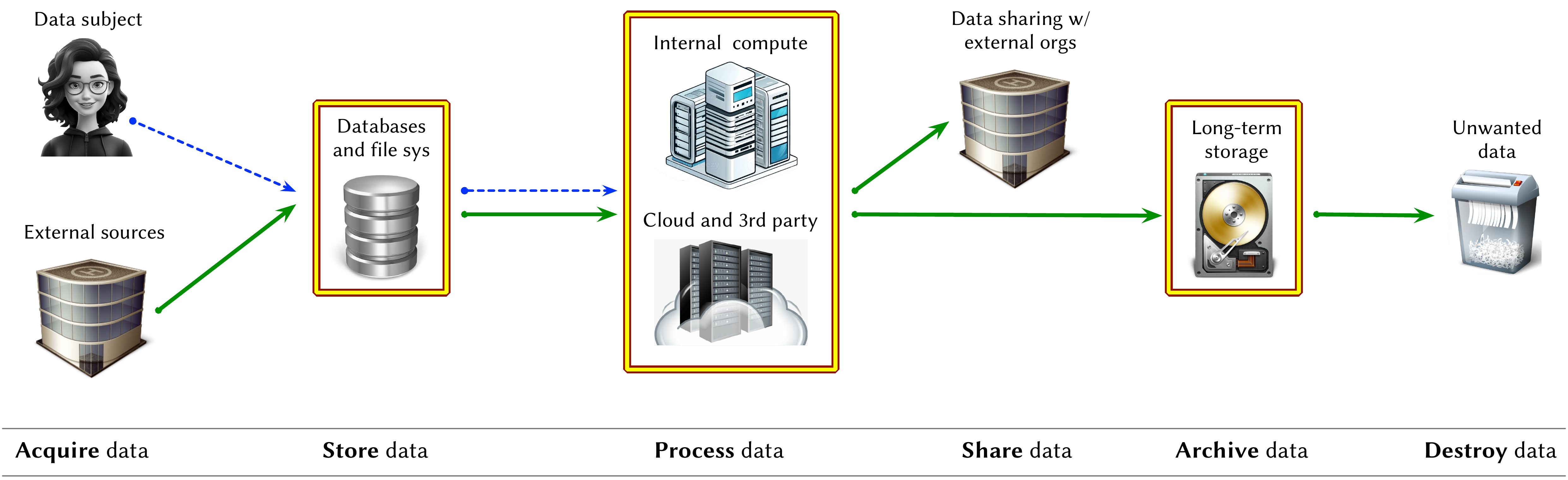}
\caption{\textbf Demonstrating the sufficiency of the RTBF tasks. Figure shows the full lifecycle of data from (left to right) acquiring, storing, processing, sharing, archiving, and to destroying. Then, we show how the four RTBF tasks (represented by blue dotted arrow, green solid arrow, and red-yellow double lines) can propagate and implement deletion at any of the lifecycle phases.}
\label{fig:rtbf-data-lifecycle}
\end{figure*}
%--------------------------------------------------------------------

\vspace{-2mm}
\subsection{Phase-1: Law-driven Design}
\label{sec:phase-1}

\vheading{Mapping Legal Intentions to Computing Tasks.}
By analyzing RTBF from a computing perspective, we identify four key tasks that cover all of its legal requirements. Figure \ref{fig:rtbf-mapping} highlights this process by showing the legal text, the computing tasks, and the mapping between the two. At this stage, it is important to keep the RTBF tasks high-level so that we do not lose the generality and interpretability afforded by the law, while at the same time, we set up a foundation to methodically explore low-level design and implementation choices in phase-2.

\vspace{1mm}
\begin{itemize}[leftmargin=2mm, parsep=2mm, topsep=1mm]
\item{\textbf{Interface with data subjects.}
This task encompasses all the input output operations of the RTBF machinery. An RTBF capable system must provide a mechanism for people to submit an RTBF request and to get a response whether the said data was forgotten (and if not, an explanation for not doing so). While \gdpr17 does not require any particular modalities for interfaces, other articles stipulate that interfaces should not impose additional burden for exercising data rights. Commonly offered interfaces for RTBF include web forms, mobile apps, email, postal mail, and telephone. In practice, the interfacing task also filters out spurious RTBF requests by authenticating the person making the request.} 

\item{\textbf{Establish deletion policies.}
RTBF is not an absolute right. For an RTBF request to be honored, it has to meet at least one of the six conditions (laid out in \gdpr17.1a through 1f), and not fall under any of the five broad exemptions (specified in \gdpr17.3a through 3e). The goal of this task is to resolve this contention and produce a yes or no decision (i.e., whether to honor an RTBF request or not). In real world, this task has proved quite challenging to be fully automated. For instance, Google Search still manages its policy resolution as a human-driven process, taking 6 days, on average, to resolve a request \cite{google-rtbf-ccs}. This task should also govern if data that is shared with external applications and datastores should also be deleted.}

\item{\textbf{Erase data from data processing systems.}
This task is responsible for erasing data from all software and hardware systems that process data including application software, libraries and system software, operating systems, hardware processors, cloud and other external processing systems. In order to make data processing performant and reliable, these systems may keep data in their in-memory data structures, in caches and runtime engines, in logs, in networked and remote processes, among others. So, the goal of this tasks is to propagate the deletion request programmatically to all subsystems (both internally and externally) and to ensure that data is deleted in a timely manner. The main challenge from a computing perspective is that many subsystems lack native support for fine-grained data deletion, and many systems implement them as lazy or shallow deletes.} 

\item{\textbf{Erase data from data storage systems.}\footnote{GDPR does not demarcate between systems that store data vs. those that process data. However, from a computing perspective, this distinction is significant. For e.g., they are different subfields of the domain. So, we explore them under two distinct tasks.}
The final task is to erase data from software and hardware systems that offer long-term, persistent storage for data. These include database systems, file systems, cloud storage, backup systems, among others. The key challenge from computing perspective is that most systems are optimized for performance, scalability, and reliability, which makes data deletion a second-class operation. This manifests in not having APIs to delete data from certain subsystems, not offering time guarantees on deletions, etc.} 

\end{itemize}

\vheading{Necessity and Sufficiency of RTBF Tasks.}
We want to establish that the four identified RTBF tasks are both necessary and sufficient to meet the legal obligations of RTBF. To ascertain the necessity of the four RTBF tasks, we employ proof by contradiction. Consider an RTBF-compliant system that does not perform \emph{interfacing with data subjects}. In such a system, no data subject can send their request to be forgotten, nor would they hear back from the system if their data was erased. This would be a violation of the legal requirements of RTBF (as stated in the first sentence of \gdpr17), in turn contradicting our assumption that the system was RTBF compliant. Next, let us consider an RTBF-compliant system that does not have \emph{established policies for deletion}. In absence of policies, such a system has to blindly honor all RTBF requests or summarily reject all RTBF requests. However, the first approach violates the legal requirement that any exercising of the individual rights must be balanced against the public interest (as stated in \gdpr17.3(c)) and the latter approach simply devoid data subjects of their RTBF rights. Thus, absence of policies would make the system non-compliant with RTBF, a contradiction. Finally, consider an RTBF-compliant system that does not have the ability to \emph{erase data from application software} or the capability to \emph{erase data from data management systems}. When such a system receives a legitimate RTBF request that must be honored, it will be unable to meet its obligation of erasing data without undue delay (as specified in \gdpr17.1), thereby invalidating the original assumption on RTBF compliance. Thus, we have established the necessity of the four RTBF tasks in achieving RTBF capability.

\vspace{1mm}
To establish sufficiency, we consider the lifecycle of data from a computing perspective. As shown in Figure \ref{fig:rtbf-data-lifecycle}, starting from the left, data is acquired either directly from data subject or from another controller; then it is stored in databases, file systems, and other short- to medium-term storage systems; third, it is consumed by data processing infrastructure; optionally, shared with external organizations; then it may be archived for long-term storage and retrieval; and finally, when it is no longer needed, it gets destroyed. Since all data have to be in one or more of these phases (by definition of the lifecycle), if we can show that the four RTBF tasks can propagate deletion request and bring about deletion in all these phases, then we would have established the sufficiency condition. First, we see that the RTBF tasks three and four (i.e., \emph{erase data from data processing systems} and \emph{erase data from data storage systems}) naturally cover the store, process, and archive phases (as marked in double yellow-red lines). Next, the RTBF task two i.e., organization's deletion policy, can specify rules for how an RTBF request is propagated from one phase to another, how each component responds to such a request, and how to handle any failures. We indicate this control flow in solid green arrows in Figure \ref{fig:rtbf-data-lifecycle}. Finally, RTBF task one, namely \emph{interface with data subject}, can propagate deletion requests from data subjects to organization's storage and compute phases, and convey their responses back to data subjects (as seen by dotted blue lines). Thus, we have established that the four RTBF tasks are sufficient to cover all the deletion-related control and data flows in the full lifecycle of data.

\subsection{Phase-2: Enforcement-driven Refinement}
\label{sec:phase-2}

Enforcement is the process of ensuring that people and organizations are complying with the law. This is typically carried out by authorized entities who (i) investigate any reported or discovered violations of the law, (ii) interpret what the law requires of that situation, and then (iii) adjudicate any punishment, correction, or fine as appropriate. Such enforcement decisions collectively identify a set of behavior that are non-compliant with the current interpretation of the law. In legal domain, this is commonly referred to as the \emph{precedent} or \emph{state of the art}. In our two-phase design process, we envision utilizing the collective knowledge from RTBF enforcements to refine the design and operation of computing systems.

\vspace{1mm}
\vheading{Tracking the Enforcement of GDPR.}
GDPR follows a distributed model of enforcement. Though the law is legislated centrally by the European parliament, its implementation is left to member nations, each of whom must enforce it within their countries. Thus, on the ground, GDPR is enforced by $\sim$28 independent and distributed agencies called the Data Protection Authorities or DPAs. Each DPA operates autonomously, determining their own priorities and enforcement strategies, and working within the budget allotted by its national government. Next, to avoid significant divergence in the application and interpretation of GDPR across countries, the law has set up a trans-national agency called the European Data Protection Board (EDPB), which can offer biding rules to DPAs on disputed issues. Finally, GDPR allows the people and organizations to challenge the enforcement decisions of DPAs and EDPB in national courts as well as the European Court of Justice (ECJ). Thus, to get a complete view of GDPR enforcement, one must procure information from all these official sources.

\vspace{1mm}
There are several projects that track the enforcement of GDPR by crawling, collecting, and annotating enforcement decisions by DPAs, courts, and EDPB. Prominent ones include GDPRhub \cite{gdprhub}, Enforcement Tracker \cite{enforcement-tracker}, and GDPRxiv \cite{sun2023gdprxiv}. We decided to use the corpus from GDPRxiv since it is based on an open methodology and also it contains a superset of decisions across all repositories. While there are 4206 enforcement decisions in the first five years of GDPR (between 25-may-2018 to 25-may-2023), we are interested in 205 of them that cite \gdpr17. Our analysis and findings in the rest of the section are based on these RTBF related decisions.

%--------------------------------------------------------------------
% Identifying legal computing tasks 
%--------------------------------------------------------------------
\begin{figure}[t]
\centering
\includegraphics[width=0.45\textwidth]{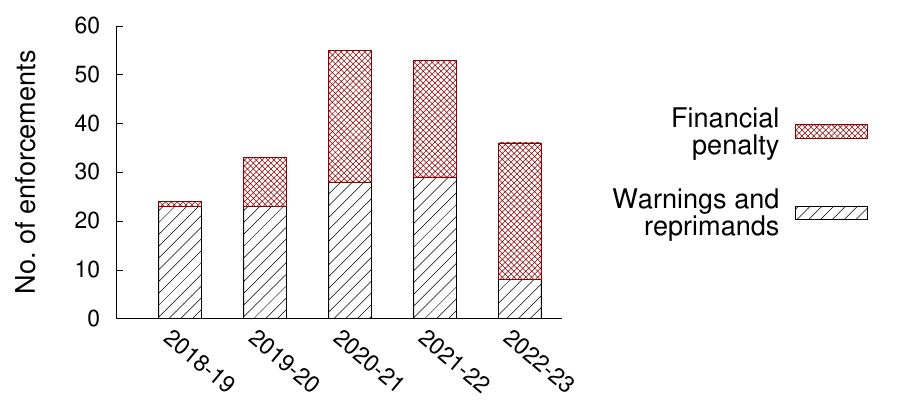}
\caption{\bf Distribution of RTBF enforcements over the years.}
\vspace{-2mm}
\label{fig:rtbf-distribution}
\end{figure}
%--------------------------------------------------------------------

\vspace{1mm}
\vheading{Understanding RTBF Enforcements in the Field.}
Figure {fig:rtbf-distribution} shows the distribution of RTBF enforcements over the five years. A simple analysis of the metadata yields many interesting observations. RTBF enforcements have been issued, on average, once every 8.9 days. About 55\% of these enforcements are warnings and criticisms that do not impose a financial penalty (these tend to be cases where the violation is a first-time offense, did not affect a large number of the people, and where the controller cooperated with the DPA to fix their violation). Amongst the cases that were levied a financial penalty, the highest was \texteuro 27.8M by the Italian DPA on the telecommunication company TIM Group and the average penalty over the five year period has been \texteuro 1.09M. Google received the most enforcements of any organization at a count of six. While these are interesting high-level observations, our goal is to analyze RTBF failures from a computing perspective, and to identify how they influence the design choices and tradeoffs for computing systems.

\vspace{1mm}
To do so, we undertake a qualitative analysis on the RTBF corpora from a computing perspective. This analysis is performed by the lead author, a PhD candidate in computer science with prior research experience in GDPR. First, by employing an open coding methodology, she extracted the main reason(s) that led to the RTBF violation. While a vast majority of them stemmed from a single reason, 18\% of them pointed to two reasons, and 1.5\% of them had three reasons for failure. Next, based on the failure reason, each of these cases is mapped to one or more of the foundational blocks of RTBF (as identified in \sref{sec:phase-1}). To validate the correctness and accuracy of the process, the other two authors of the paper (a law scholar and a computer scientist, both with GDPR experience), randomly picked 10\% of documents to carry out coding and mapping independently. We did not find any meaningful discrepancies or divergences between the two analyses. Though this \emph{catalog of failures} is a useful reference for RTBF designers, the annotations are too big to be included in the paper verbatim, even in the appendix, so we make them available on the project website. However, the failures exhibit several recurring themes and patterns, which render them to be studied in logical groups of RTBF properties they violate. Below, we present eight of the most commonly occurring themes: 

\vspace{2mm}
\begin{itemize}[leftmargin=6mm, parsep=2mm, topsep=1mm]
\item{\textbf{Exercising RTBF.}
Regulators have issued several sanctions for making it harder for people to exercise RTBF. In the first category (that we call \emph{UI-TROUBLE}), we group eight cases where controllers did not specify how to submit RTBF requests, pointed to RTBF webpages that errored out, or provided RTBF emails that were undeliverable, among others. In one case, the Spanish regulator fined Google for making people fill out a pre-questionnaire before allowing them to submit RTBF requests. The second broad category (that we refer to as \emph{VERIFY-MORE}) includes eight cases where controllers employed stringent verificiations before allowing RTBF. For instance, an Austrian online service required a physical signature; Groupon and Twitter required a national ID. These practices were deemed disproportional since such verifications were not required when creating accounts or using their services.} 

\item{\textbf{Response Time.}
It measures the time between making an RTBF request and receiving the first response from the controller. A recommended value comes from \gdpr12(3), which states that all data right requests should be responded within one month of receipt (while allowing exemptions for cases requiring complex processing). In 54 cases, the controller never responded (we refer to this category as \emph{NO-RESPONSE}) and in 25 cases, the controller took more than 30 days to respond (category \emph{LATE-RESPONSE}). These enforcements have clarified the minimum expectation: controllers must acknowledge RTBF requests within a month of receipt; though RTBF decision-making and actual data deletion may take longer.}

\item{\textbf{Explainability.}
This represents the ability of a data subject to understand how their RTBF request was handled. There are eight cases in this category (we denote them by \emph{EXPLAIN}), all of which made it clear that explanations are important when rejecting RTBF requests. For instance, in two separate cases against Google and Spotify, regulators agreed that the companies were correct in rejecting the RTBF requests but fined them for lacking proper explanation in their responses. Secondly, precedents indicate that the controllers must explain the key reasoning behind their decision. This could be due to influence of other laws (for instance, RTBF cannot be exercised on betting data since online betting laws require this data to be kept for five years---Bet365 in Denmark); legal obligations (say, the controller keeps a log of all RTBF requests and no RTBF will be supported on that log itself---unnamed company in Belgium); or business practices (such as not allowing RTBF on payment default data unless the debt has been paid off---unnamed company in Hungary). Straight forward as these are, none of these explanations were included in the RTBF responses, and all three controllers were fined or warned for omitting them. So, the prudent option when rejecting RTBF is to include a precise and individualized explanation.}

\item{\textbf{Exemption handling.}
This is the process of vetting if an RTBF request meets all the legal criteria. Accommodating exemptions has proven to be a difficult task in the field. Thus far, 41 cases say exemptions from within GDPR were misinterpreted (we denote this as \emph{EXMPT-GDPR}), 24 cases say the interplay of RTBF with non-GDPR laws was disregarded (category \emph{EXMPT-OTH}), and 13 cases say both were incorrectly handled (category \emph{EXMPT-ALL}). The complexity of this process is reflected in Google search's approach of using skilled human arbiters who take up to 6 days on average to arrive at an RTBF decision \cite{google-rtbf-ccs}. Despite those guard rails, Google has been fined four times for incorrect exemption handling. Interestingly, we have not seen a case where a controller was penalized for honoring an RTBF request that should have been rejected. So, it may be prudent to weigh more in favor of honoring the RTBF request unless the applicability of exemptions is glaringly obvious.}

\item{\textbf{Deletion propagation.}
This is the process of conveying the delete request to all software and hardware systems that have copies of the RTBF data, and ensuring that the data is indeed deleted. In this category (\emph{PROPAGATE}), five enforcements have highlighted shortcomings in both internal and external propagations. For example, the Swedish regulator reprimanded Rebtel Networks for continuing to send marketing emails even after an RTBF on email address was granted--a failure to propagate deletion across all internal data processing/storage systems. On the external propagation side, the Danish regulator issued a criticism against Høje-Taastrup municipality for using a third-party processor who did not offer APIs for deletion on demand.}

\item{\textbf{Deletion APIs.}
This category (denoted by \emph{NO\_API}) comprises of 15 cases where data processing systems did not support deletion well. The issue manifested in several forms: organizations relying on employees to manually enter delete requests to multiple disjoint systems, which led to errors; software bugs that resulted in delete requests being silently dropped before processing; setting of incorrect values for the expiry time of the data; companies switching to new software systems which could not make deletions on older data; organizations using third-party processors who did not offer deletion APIs; an AI company unable to detect all the data items in their training set that matched the RTBF criterion; and a government organization that did not have administrative privileges to delete data from their web system.}

\item{\textbf{Service degradation.}
Enforcements in this category represent cases where the controller refused RTBF because honoring it would result in service degradation or no service to the customer. There are five examples in this category (denoted by \emph{BAD\_SERV}) including: a company in Austria that could not offer loyalty program benefits to people who request RTBF on email address and phone number; Taxa in Denmark would erase all prior transactions of the customer if they exercised RTBF on phone number, in turn degrading their future service. However, contrastingly, Spotify was allowed to reject RTBF request on credit card information for those customers who used free trials in order to prevent future abuse of the program.}

\item{\textbf{Database constraints.}
This captures issues stemming from database and storage systems that prevent (or make it difficult in) implementing RTBF. There are three enforcements in this category (that we denote by \emph{DBMS}) including: Carrefour in France couldn't delete email addresses and Taxa in Denmark couldn't remove phone numbers due to primary key and foreign key constraints in their database schema. In all these cases, the regulators deemed it an insufficient reason for rejecting RTBF requests and in turn, ordered them to re-architect their data management systems.}

\item{\textbf{Other frequent decisions.}
We observe a number of enforcement decisions that likely stem from incompetence or ignorance of the controller (that are nonetheless repeated frequently). For example, in 17 cases, the controllers responded falsely that RTBF was completed when in reality they took no such action internally (category \emph{OMIT}); in three cases, the controllers tried to avoid RTBF responsibility by redirecting people to another controller (category \emph{DEFLECT}); lastly, in two cases, the controllers could not authenticate the requester as the data was obtained by legacy/unknown processes (category \emph{NO-SRC}).}

\item{\textbf{Decisions without a category.}
Finally, we are left with five decisions that neither fit into an already named category nor have a frequency of more than one. While this does not make them any less important (in terms of setting a precedent), we have chosen to leave these one-off decisions uncategorized. These include a Danish case where the controller initially refused RTBF resulting in a complaint by the data subject but when the Danish regulator started the investigation, they found that data was already deleted; a Hungarian case where a law firm requested RTBF on their name being included in a negative list but the regulator reminded them that RTBF only applies to humans; and three other cases.}

\end{itemize}

%--------------------------------------------------------------------
% Enforcement categories
%--------------------------------------------------------------------
\begin{figure}[t]
\centering
\includegraphics[width=0.5\textwidth]{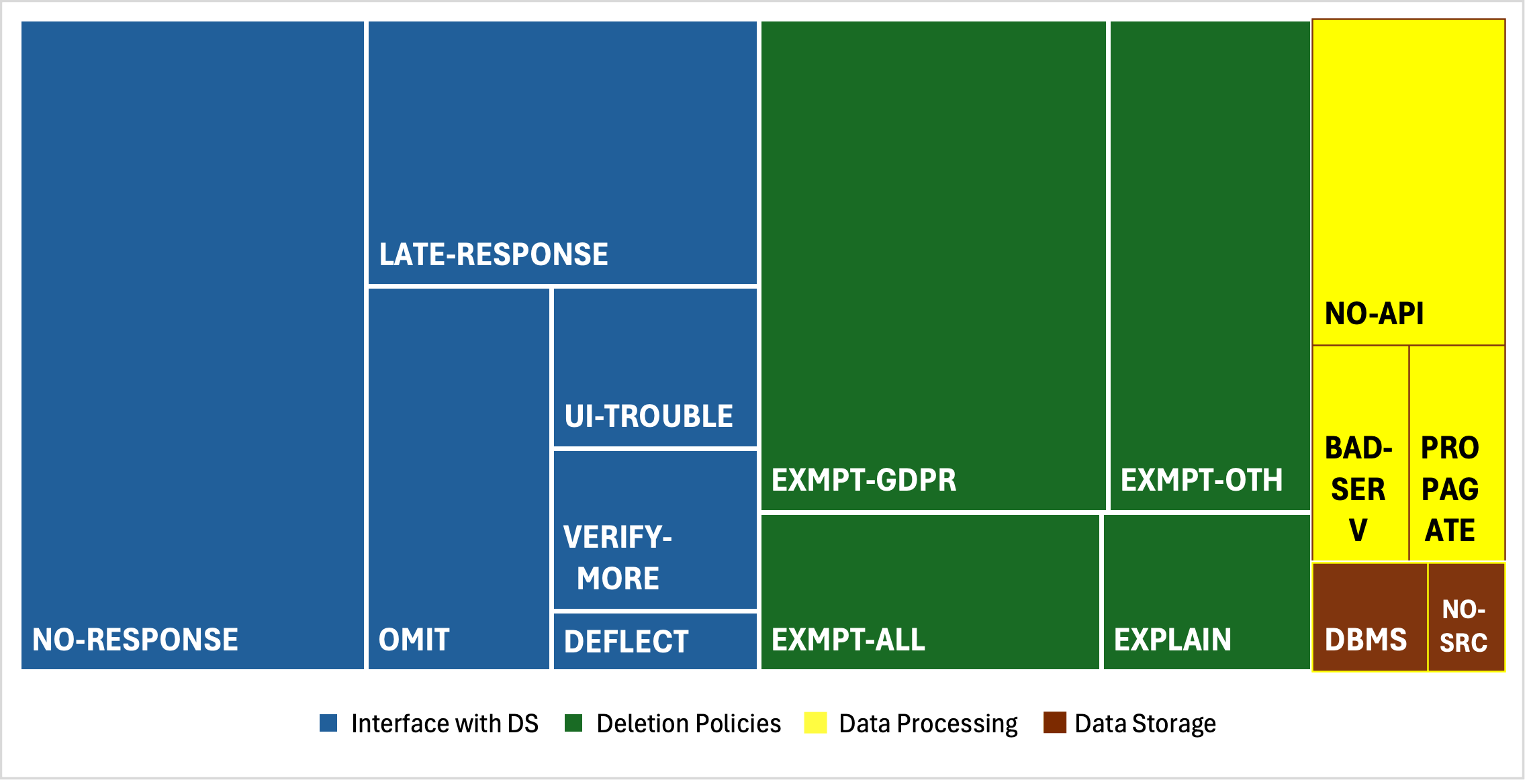}
\caption{\textbf {A treemap of RTBF enforcements, where each box represents a named category. The area of the box is proportional to the number of cases in that category and the color of the box indicates which of the four RTBF tasks it failed.}}
\vspace{-2mm}
\label{fig:enf-categories}
\end{figure}
%--------------------------------------------------------------------

\vspace{1mm}
\vheading{Visualizing RTBF Enforcements.}
To better understand the distribution of RTBF enforcements, we plot a treemap in Figure \ref{fig:enf-categories}. Each rectangular box represents a named category with its area being proportional to the number of cases within of that category. In addition, we color the boxes to show which of the four RTBF tasks they failed. This helps us recognize two trends: first, we see that 49\% of enforcements cite failures in interfacing with data subjects, and 37\% cite failures in policy resolution. While this distribution is skewed, it is not entirely surprising since UI and policy failures are easier to catch and they impact people immediately (as opposed to failures in data processing and data storage systems). Next, we see that more than half the failures are rooted in just three categories (\emph{NO-RESPONSE, LATE-RESPONSE, EXMPT-GDPR}). Understanding these macro trends could help organizations reevaluate their RTBF activities and focus on areas of high enforcement.

\subsection{Scope and Limitations}
\label{sec:design-limitations}
\vspace{1mm}

The proposed technique of \emph{design based on law, and refine based on enforcements}, while effective in creating an RTBF infrastructure and in avoiding known failures, cannot ensure 100\% compliance. So, it is important to discuss the limitations and scope. First, \emph{compliance is contextual} i.e., what an organization must do to be compliant is dependent on a variety of factors such as type and volume of personal data, nature of processing, security measures employed, risks to the people, size of the organization, among others. Thus, a compliance profile for Wikipedia is going to be different from Google Search or a children's hospital. Second, \emph{compliance evolves with time} i.e., in order to stay compliant, an organization must continuously evaluate their design choices and operating practices to ensure consistency with new precedents. Thus, phase-2 is never complete but instead be treated as a periodic health check activity for systems and processes. Lastly, \emph{compliance is costly} i.e., it makes organizations do things that they may not have done otherwise. So, it may be natural for organizations to find ways to minimize compliance costs. However, our technique is not focused on optimizing for cost or performance, nor can it be readily automated. 

%TBA: cases that were upheld or rejected are not included in our analysis.

\section{Evaluation}
\label{sec:evaluation}

We evaluate the usefulness of our technique by answering two questions: (1) Does it help avoid RTBF violations in the future, and (2) Could it identify RTBF pitfalls stemming from past practices?

\subsection{Avoiding Faults in the Future}
\label{sec:future-rtbf-violations}

To determine how capable our system is in avoiding or reducing future violations of RTBF, we undertake a longitudinal study. We ground this in real-world by structuring it as a retrospective investigation i.e., does the knowledge from first five years of RTBF help avoid violations that occurred in year-6 of GDPR.

\vheading{Experiment setup.}
Since it has been more than a year that we started this project, enforcements from year-6 are already available on GDPRxiv. So, we gather all enforcements of year-6 (i.e., those between May 25, 2023 and May 25, 2024) that cite \gdpr17. First, we annotate these 40 enforcements following the same methodology as outlined in \sref{sec:phase-2}. Next, we compare each of these annotations to the annotated corpora from the first five years. To do that, we employ IRAC (short for \emph{Issue-Rule-Application-Conclusion}, a method practiced by legal scholars in analyzing cases). The key advatage of this method is that splits the analysis into four distinct parts: (i) the \emph{Issue} part identifies the legal question or the dispute that needs to be resolved, (ii) the \emph{Rule} portion lists all the applicable rules for the case, (iii) the \emph{Application} section consists of applying the identified rules to the given situation, and (iv) the \emph{Conclusion} outlines the final decision. The first and the fourth parts of all our cases are similar since they capture enforced failures in RTBF. However, the middle two aspects help us define a mapping function as follows: 

\vspace{1mm}
\begin{center}
\noindent\fcolorbox{black}{white}{\parbox{0.95\linewidth}{
\textsc{\textbf{strong\_match(e, c)}}: an enforcement \textsc{\textbf{e}} has a strong match to category \textsc{\textbf{c}}, if its annotation points to \textsc{\textbf{c}} as a failure category (\emph{Rule} match) and there exists an enforcement \textsc{\textbf{e'}} in \textsc{\textbf{c}} that has the same application of that specific rule (\emph{Application} match). This mapping indicates that if the controller had known about \textsc{\textbf{e'}} and fixed the specific RTBF condition that led to a failure in \textsc{\textbf{e'}}, they could have avoided \textsc{\textbf{e}}.
\\[2mm]
\textsc{\textbf{weak\_match(e, c)}}: an enforcement \textsc{\textbf{e}} has a weak match to category \textsc{\textbf{c}}, if its annotation points to \textsc{\textbf{c}} as a failure category (\emph{Rule} match) but there is no enforcement \textsc{\textbf{e'}} in \textsc{\textbf{c}} that has the same application of the specific rule (no \emph{Application} match). This mapping indicates that if the controller had known about category \textsc{\textbf{c}} and its constituent cases, there is a reasonable chance to avoiding \textsc{\textbf{e}}.
\\[2mm]
\textsc{\textbf{no\_match(e, c)}}: an enforcement \textsc{\textbf{e}} has no match to category \textsc{\textbf{c}}, if its annotation does not indicate \textsc{\textbf{c}} as a failure category (no \emph{Rule} match) and by extension, no \emph{Application} match. This mapping indicates that even if the controller had known about category \textsc{\textbf{c}} and its constituent cases, they gain no knowledge that helps them avoid \textsc{\textbf{e}}.
}}
\end{center}
\vspace{1mm}

\begin{figure}[t]
\centering
\includegraphics[width=0.5\textwidth]{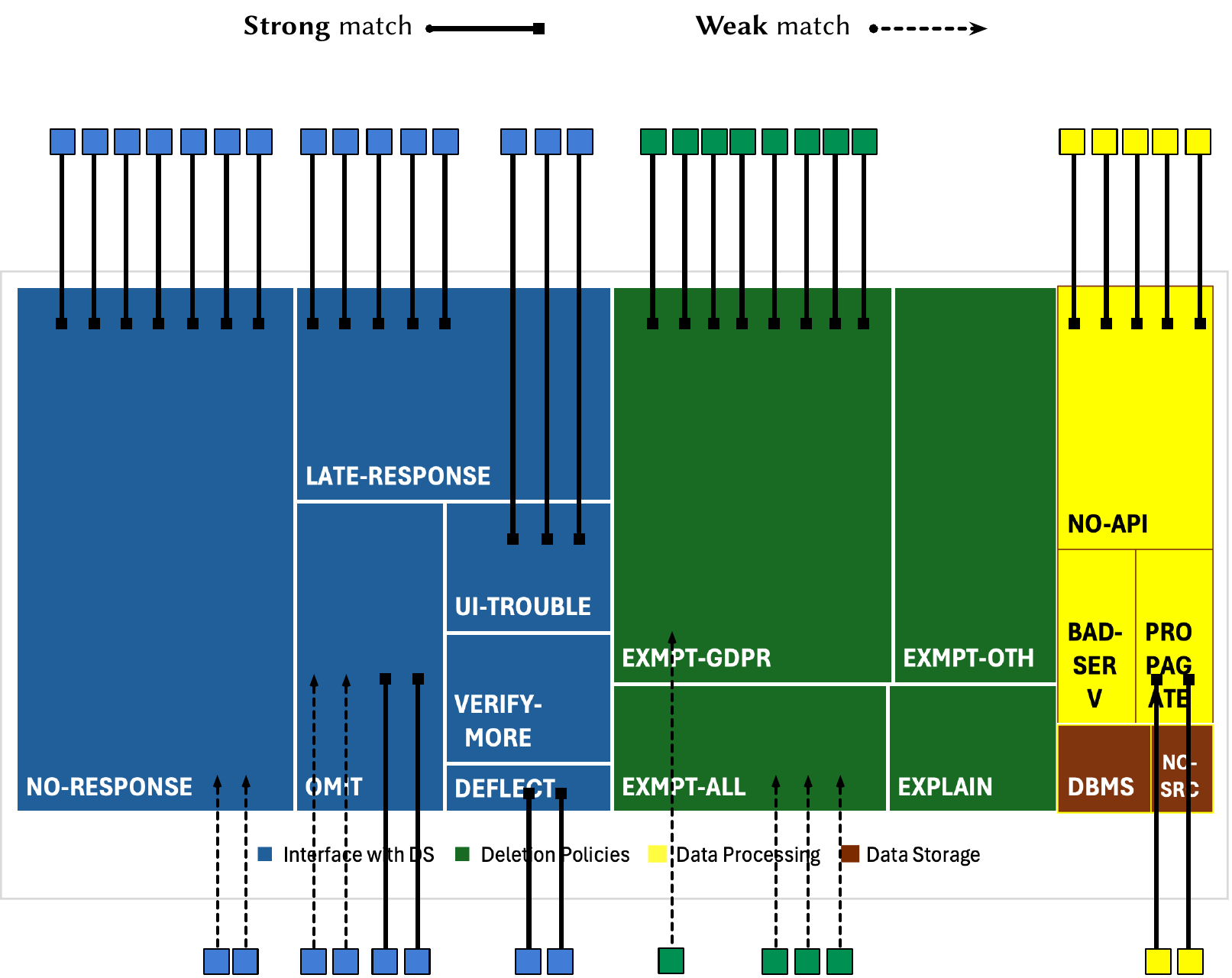}
\caption{\textbf {Matching year-6 enforcements}}
\label{fig:enf-mapping}
\vspace{-4mm}
\end{figure}

Then, we apply this mapping function on all of the year-6 cases, comparing them with all the named categories and their constituent cases. A \textsc{\textbf{no\_match}} is easy to determine since we only have 15 named categories. For cases that do match a category, we iterate until a \textsc{\textbf{strong\_match}} is found or until the iterations terminate having run out of enforcement cases in all its matched categories, at which point, we note it down as a \textsc{\textbf{weak\_match}}. A complete mapping of all year-6 cases along with our justification for the mapping choices is available on the project website. However, Figure \ref{fig:enf-mapping} represents this pictorially with each case being represented by a rectangular point, and mappings indicated by lines (thicker lines for strong mapping and dotted lines for weak mapping).

\vspace{1mm}
First off, we see that none of the 40 enforcements of year-6 created any new categories. 32 cases exhibited \textsc{\textbf{strong\_match}} while the remaining eight were \textsc{\textbf{weak\_match}}es. This observation strongly supports our hypothesis that learning about prior RTBF enforcements and refining one's design and operation is an effective way in avoiding future RTBF failures. The figure also shows how year-6 enforcements are distributed differently compared to prior enforcements. For example, we see a reduction in exemption handling, explainability, and verification categories; whereas an increase in the categories of no\_api and UI trouble. This is an indication that future enforcement priorities may vary over time, and evolve with the field. Though it did not happen in year-6, it is quite possible that new failure categories are recognized, especially with the emergence of new technologies and as controllers avoid making commonly identified mistakes.

%--------------------------------------------------------------------
% Enforcement categories
%--------------------------------------------------------------------
%--------------------------------------------------------------------

%\vspace{-2mm}
\subsection{Avoiding Pitfalls from the Past}
\label{sec:RTBF-anti-patterns}

We apply learnings from the two-phase technique to re-evaluate certain long-standing practices of computing and data management systems that create pitfalls for RTBF compliance. This exploration is inspired by the notion of anti-patterns, which were originally conceived by Andrew Koenig \cite{koenig-antipattern} and later expanded by Brown et. al., \cite{antipatterns-book}. They define anti-patterns to be the principles and practices of software design that seem intuitive and useful in a narrow context, but prove ineffective in a broader system and over long term. Examples of Koenig's anti-patterns include (i) premature optimization of individual functions in a large software system, (ii) use of magic numbers in programming, and (iii) creating a god class in object design, among others. 

\vspace{1mm}
In a similar vein, we define \emph{RTBF anti-patterns} as long-standing principles and practices of computing systems that serve their original purpose well, but make it harder to support RTBF. After analyzing the RTBF enforcement corpora, we have discovered six such RTBF anti-patterns that we present below.

\vspace{1mm}
\vheading{Scope and limitations.}
The anti-patterns presented here should be treated as starting points towards making systems adapt better to RTBF. They should not be considered \emph{exhaustive} (i.e., it may be possible to identify additional anti-patterns by analyzing the corpora with a different perspective) or \emph{prescriptive} (i.e., there is no guarantee that having these in your systems will incur a penalty nor that avoiding them will be enough to avoid a violation). Finally, anti-patterns are distinct from dark patterns \cite{dark-patterns-1, dark-patterns-2, dark-patterns-3, dark-patterns-4, dark-patterns-5} which are techniques and practices aimed at deceiving the users and manipulating them into taking certain actions. In contrast, anti-patterns do not have any malicious intent built-in; instead, they emerge by looking differently at principles and practices that have served as good patterns elsewhere.

\vspace{4mm}
\noindent\fcolorbox{black}{gray!20}{\parbox{0.98\linewidth}{
\textbf{ 1. Using personal data as primary keys}
}}
\vspace{1mm}

\noindent Personal data, especially those that uniquely identify people, say phone numbers, email, or national IDs, have long been used as primary keys in database systems. While GDPR does not preclude this practice, it often results in practical difficulties in honoring RTBF. For instance, when the Danish regulators found that Taxa (a Copenhagen based taxi company) had difficulties in removing user data because of database systems constraint, they were fined and asked to redesign their data storage systems. Similarly, Carrefour (a French retail company) resorted to using ad hoc tools to circumvent database system constraints during an investigation by the French regulator, ultimately resulting in \texteuro 2.25M penalty.  

\vspace{4mm}
\noindent\fcolorbox{black}{gray!20}{\parbox{0.98\linewidth}{
\textbf{ 2. Anonymizing personal data instead of deleting}
}}
\vspace{1mm}

\noindent Organizations may choose to anonymize the personal data upon getting an RTBF request \cite{irish-dpa-anonymization, wp-anonymization, edps-anonymization}. This is appealing because it lets the controller derive some benefit from the data that otherwise would be permanently deleted. However, a robust anonymization is hard to achieve. For example, in 2019, the Danish DPA fined a taxi company for anonymizing its 9 million taxi ride records while retaining enough auxiliary data to be able to reconstruct the dataset \cite{danish-taxa-rtbf}. Even when it is well done, prior research has shown that not all anonymizations hold the test of time \cite{de-anonymize-1, de-anonymize-2} and may need to be constantly evaluated to ensure anonymity \cite{de-anonymize-3}. Lastly, it is important to note that GDPR considers the act of \emph{anonymizing personal data} as a data processing activity in itself \cite{edps-anonymization}. So, the controller must ensure that they have a legitimate purpose or a legal basis to perform anonymization and it cannot be used as a tool to evade the applicability of GDPR. A case in point: in 2020, the Greek regulators reprimanded the insurance company Axa for anonymizing its customer data so that they could retain it beyond the date of expiry or date of RTBF request \cite{greece-axa-rtbf}.

\vspace{4mm}
\noindent\fcolorbox{black}{gray!20}{\parbox{0.98\linewidth}{
\textbf{ 3. Keeping personal data untagged}
}}
\vspace{1mm}

\noindent Tagging personal data with attributes such as its purpose, time-to-live, associated person, objections to its use, etc., makes it possible to comply with people exercising their data rights. However, data tagging is a resource-intensive operation, both computationally and on the storage system \cite{gdpr-vldb}. Thus, there is a tendency to avoid it or at least, delay it to a point when data subjects actually exercise their rights. For example, consider Clearview AI---a company that collects images of people's faces from the public Internet and social media to build an online global database, which is then used to offer facial-recognition-as-a-service to law enforcement and private organizations. When EU citizens approached Clearview AI to exercise RTBF, they were instructed to provide photographs instead of just their identities, so that their photos could be matched in the database. Several regulators \cite{clearview-1, clearview-2, clearview-3} found this practice to be an impediment to people exercising their RTBF rights, among other things, and issued penalties of over \texteuro40M.

\vspace{4mm}
\noindent\fcolorbox{black}{gray!20}{\parbox{0.98\linewidth}{
\textbf{ 4. Mismanaging the depth of deletion}
}}
\vspace{1mm}

\noindent \gdpr17 does not specify how deep should a deletion be percolated, instead leaving its interpretation to data controllers and regulators. So, simply extending the existing delete workflows to RTBF may lead to non-compliant behavior. At one extreme, Brico Prive, a French online retailer, had the practice of simply deactivating the user handles upon receiving RTBF requests while still retaining the user data in the database. The French regulators deemed this an \emph{incomplete deletion} and fined them \texteuro500K \cite{french-brico-rtbf}. At the other end, consider Google Search's implementation of RTBF. When they determine a link has to be delisted, they not only delete it from their databases but also propagate the deletion request to the original website that published the content. The Swedish regulators fined Google for this practice \texteuro7M \cite{swedish-google-rtbf} citing that Google had no legal basis for \emph{propagating deletion} requests externally (which would violate user's privacy).

\vspace{4mm}
\noindent\fcolorbox{black}{gray!20}{\parbox{0.98\linewidth}{
\textbf{ 5. Logging without checks and bounds}
}}
\vspace{1mm}

\noindent Logs are relatively inexpensive to generate and to store, and have been used for a variety of purposes including debugging and traceability, post-hoc analysis and audit, and as reliability primitives. Since GDPR allows (via \gdpr24 and \gdpr30) the use of computer logs as evidence in establishing compliance, organizations have tended to \emph{play it safe} in storing any and all forms of logs for possible future use. However, logging can conflict with the spirit of RTBF. In 2022, the Belgian regulators reprimanded a controller for storing excessive information about RTBF requests in their logs. Given that GDPR defines the act of logging as a data processing activity in itself, care should be taken (i) to not store any RTBF deleted data in logs, (ii) to follow the principle of data minimization i.e., do not store anything that is not essential, and (iii) to set a date for expiring old logs. Any logging system that ignores these requirements risks violating RTBF.

\vspace{4mm}
\noindent\fcolorbox{black}{gray!20}{\parbox{0.98\linewidth}{
\textbf{ 6. Employing excessive verification}
}}
\vspace{1mm}

\noindent In contrast to other GDPR rights, RTBF, by definition, is for single and definitive use. That is, once RTBF is exercised successfully (i.e., a dataset has been erased), no other right could be exercised on the same dataset. This has prompted some organizations to employ stricter and excessive forms of user verification. However, this practice was deemed as placing an undue burden on people in exercising their RTBF. For instance, the Irish regulators issued reprimands to Groupon in 2020 \cite{ireland-groupon-rtbf} and to Twitter in 2022 \cite{ireland-twitter-rtbf} for requiring national photo IDs to exercise RTBF but allowing people create accounts without such IDs. A prudent approach seems to be having a verification scheme that is uniform across all types of data interactions including exercising of data rights.

\section{Applying Our Technique in Practice}
Having presented our technique and evaluated its effectiveness in avoiding RTBF failures, we now describe how to apply it in practice. Any effort at achieving compliance must be undertaken at the organization level since GDPR applies to organizations (not to people or processes or software systems within that organization). However, it is important that all components of the organization must be individually \emph{RTBF capable}, so they can enable the organization to achieve \emph{RTBF compliance} as a whole. For instance, if an organization uses a software system that does not have the ability to delete its data, then it is almost impossible for the organization to be fully RTBF compliant. 

\vspace{1mm}
So, we recommend starting with the complete set of all software and hardware systems present within the organization, no matter their designated functionality. For a small organization with a simple data flow, it could just be their web stack or a cloud-based software service, whereas for a large organization with complex data ecosystem, it could be tens or hundreds of software and hardware systems. Next, we map each of these systems to one or more of the four RTBF tasks identified in phase-1, based on their role in the data lifecycle. This step ensures that there is at least one system responsible for each of the core RTBF tasks (if not, the controller needs to fix it). Then, for each of these mappings, we apply all the phase-2 refinements i.e., for each system performing a given RTBF task, we examine the validity of its design choices and tradeoffs as well as weed out any non-compliant behavior. Once component level capabilities are assessed, we zoom out to the system architecture level, where we can track the inter-component interactions and flow of data holistically. For instance, we should ensure that an internal system that uses an external cloud backup is invoking the correct deletion API in a timely manner to remove RTBF data from the cloud. In summary, this hierarchical approach ensures that individual systems are themselves \emph{RTBF capable}, and the overall system architecture is \emph{RTBF compliant} at the organization level.

%To provide an outline for how a controller may adapt our technique, we use Wikimedia Foundation (the organization that runs the Wikipedia service) as a reference. This choice driven by three factors. First, the data traffic and popularity of Wikipedia. It is one of the world's top-10 most visited websites with more than 1.5B unique monthly visitors and 15M monthly content edits. Second, it has an advanced data ecosystem, both in terms of data size (for example, the English Wikipedia has 62M pages and a size of 26TB) and the complexity of technical infrastructure (XXX disntinct software systems). Lastly, their transparency in organizational matter. For instance, they not only publish their software architecture but also have detailed reports on RTBF requests dating back to 2014.

\vspace{1mm}
\vheading{Scope and limitations.}
Our primary focus is to explore RTBF from a computing perspective. So, we exclude non-computational components of the organization such as employees and processes as well as traditional pen-and-paper data processing systems. Also, we must emphasize that our method does not guarantee 100\% RTBF compliance, nor should it be considered universally applicable. Instead, our approach should be treated as \emph{one of the tools} available for data controllers in methodically analyzing the RTBF requirements and in rooting out known failures. We have further clarified our scope and limitations in \sref{sec:design-limitations} and \sref{sec:RTBF-anti-patterns}.

\subsection{RTBF Capable Elasticsearch}
\label{sec:elasticsearch}

To demonstrate our approach concretely, we select search engine as reference system. This choice is driven by several factors. First, given the importance of searching in computing, search engines are ubiquitous in software stacks. Second, search engines comprise of both data processing and data storage functionalities of RTBF, thereby providing a broader context for our technique. Finally, search engines were the first software system to be subject to RTBF (more details in \sref{sec:rtbf-primer}). Importance of RTBF in search engines is emphasized by the fact that EDPB published a dedicated guidance on this matter in 2020 \cite{edpb-rtbf-search-engines}. Our analysis and evaluation in the rest of the section are based on Elasticsearch \cite{elasticsearch}, an open-source search engine that is amongst the most widely deployed in the world.

\subsubsection{\textbf{Is Elasticsearch RTBF Capable? }}
\hfill
\vspace{1mm}

\noindent Elasticsearch is a distributed search engine that can support search analytics in near real-time at petabyte scale. Given its prominence in the data analytics ecosystem, several prior work have analyzed its GDPR readiness \cite{elastic-gdpr, elastic-gdpr-2}. However, in these analyses, support for RTBF has been simply reduced to \emph{whether Elasticsearch provides APIs to delete data}. We aim to explore this question in more depth and with nuance, based on our two-phase design. Since Elasticsearch is primarily a data processing and data storage system, our exploration will focus on these (while covering any applicable aspects of UI and policy layers). Below, we investigate Elasticsearch's RTBF capability along two axes: what RTBF tasks can it perform and how well does it perform them:

%--------------------------------------------------------------------
% Elasticsearch components with data 
%--------------------------------------------------------------------
\begin{table*}[t]
\caption{\bf Exploring the component-level data storage in Elasticsearch}
\label{tbl-cleansing-delete}
\makebox[1\textwidth][c]{
\begin{minipage}[b]{1\textwidth}
  \centering
  \setlength{\tabcolsep}{8pt}
  \renewcommand{\arraystretch}{1.2}
  \begin{tabular}{@{}l | l | l | l | l | l | l@{}}
    \toprule[1.2pt]
    \midrule
    & \makecell[l]{Data structures} & \makecell[l]{Elastic caches} & \makecell[l]{Translog} & \makecell[l]{Snapshots} & \makecell[l]{Event log} & \makecell[l]{Replica shard} \\[3mm]
    \toprule[1.2pt]
    \textbf{Purpose} & performance & performance & fault tolerance & fault tolerance & monitoring & availability \\
    \midrule
    \textbf{Storage location} & memory, disk & memory, cache & disk & disk & disk & external server \\
    \midrule
    \textbf{\makecell[l]{Retention beyond\\ deletion}} & until merge & \makecell[l]{depends on\\ system load} & until flush & no limit & no limit & \makecell[l]{same as\\ primary} \\
    \midrule
    \textbf{\makecell[l]{API to delete\\ select data}} & \makecell[l]{yes but timing is\\ not guaranteed} & no & no & no & no & no \\[2mm]
    \bottomrule[1.2pt]
  \end{tabular}
\end{minipage}}
\end{table*}
%--------------------------------------------------------------------

\vspace{1mm}
\begin{itemize}[leftmargin=6mm, parsep=2mm, topsep=1mm]
\item{\textbf{Deletion APIs.}
Elasticsearch offers two direct options: the RESTful method \texttt{DELETE} and the API \texttt{\_delete\_by\_query}. The former can be used to delete a document (equivalent of tuples in DBMS) or an index (equivalent of an entire database in DBMS), while the latter is useful for deleting a set of documents that match a query. Elasticsearch also provides a mechanism to remove contents within a document (equivalent of setting a field to null in DBMS) through the \texttt{\_update\_by\_query} API. Finally, Elasticsearch used to support a \texttt{time-to-live} field for each document, which ensured that a given document is automatically deleted after the specified time, but this has been deprecated since version 5 (circa June 2018).}

\item{\textbf{Deletion granularity.}
It is straightforward to see that Elasticsearch's built-in deletion APIs and methods (discussed above) allow for data to be deleted at all granularities: individual key-value pairs and documents, groups of key-value pairs and documents that match a search criteria, and the entire indexes.} 

\item{\textbf{Deletion latency.}
Elasticsearch takes a lazy approach to deletion i.e., all deletion APIs simply \emph{mark the data as deleted} without actually erasing them from data structures and files; instead, the actual deletion is carried out at a later time depending on the runtime state of the system and user-defined configuration parameters (we elaborate on this in \sref{sec:cleansing-delete}). While this approach helps Elasticsearch be highly performant, it makes it hard to determine the latency of deletion operations.}

\item{\textbf{Deletion propagation.}
Elasticsearch's deletion can be best categorized as \emph{shallow}. Deletion APIs make the deleted data unavailable for the application, and eventually remove them from the internal data structures. However, the deletion request is not actively propagated to all the underlying subsystems which may contain the RTBF data. For instance, query cache, translog, and snapshots, to name a few. So, it is fair to say Elasticsearch does not have robust internal propagation of deletion.}
\end{itemize}

\noindent In summary, Elasticsearch has basic RTBF capability but falls short on two aspects: it does not completely erase the RTBF data from all underlying subsystems, and even for those subsystems that it does, it is hard to estimate when the actual erasure happens. These behavior, while not unique to Elasticsearch \cite{lazy-delete-1, lazy-delete-2, boston-deletion}, do conflict with findings from four enforcements in \emph{NO\_API} and \emph{PROPAGATE} categories. Our next subsection proposes a resolution to this.

\vspace{1mm}
\subsubsection{\textbf{Timed and Thorough Deletion in Elasticsearch}}
\hfill
\label{sec:cleansing-delete}

\vspace{1mm}
\noindent We describe a deletion operation as \emph{timed} if it has a notion of time associated with it. For example, TTL-based deletion, where data would be deleted at a predetermined time; or real-time deletion, where data would be deleted synchronously in real-time. Similarly, we characterize a deletion to be \emph{thorough} if all copies of the given data are erased from all parts of the system. This is distinct from sanitizing deletes \cite{secure-deletion-disk} where data has to be irrevocably erased from the physical hardware; instead, a thorough delete simply goes through all the subsystems that may contain the given data and issues a delete request on them. Combining these two properties, we define \emph{cleansing delete} as a delete operation that is both thorough and timed.

\vspace{1mm}
Despite being grounded in RTBF enforcements, a cleansing delete may seem unnecessary or excessive. However, if one were to consider deletion as a first-class operation, then optimizing for its quality and clarity would seem like a natural choice. We think of this as extending the notion of \emph{clean up after yourself} to data management systems. In fact, cleansing deletes are not without a precedent. Consider, Google cloud's deletion guarantee \cite{google-cloud-deletion} that all copies of the data would be deleted from all underlying subsystems within 180 days of requesting it. In contrast, Elasticsearch's deletion is neither thorough nor timed. Our goal is to investigate why Elasticsearch’s deletion is not cleansing, how to fix it, and if it impacts the performance.

\vspace{1mm}
\vheading{Elasticsearch's data footprint.} Table \ref{tbl-cleansing-delete} shows the main components of Elasticsearch in which data gets stored during the create, read, and update operations. First, the internal data structures. Both Elasticsearch and Lucene (the search engine library on which Elasticsearch is built) employ a variety of data structures to efficiently index, query, and manage documents. These include inverted index, k-dimensional B-Trees, document-value structures, deleted document index, among others. Depending on the size of data, access patterns, available hardware resources, etc, these data structures will be stored in either in memory or on disk (or split between the two). While Elasticsearch delete APIs allow fine-grained deletions, they simply mark the deleted item with a \texttt{tombstone} but leave them as is in the data structures. So, the only way to guarantee a complete and immediate removal of deleted data is to trigger a \texttt{forcemerge}---a practice not recommended by Elasticsearch \cite{elasticsearch-forced-merge}.

\vspace{1mm}
Second is the caching system used by Elasticsearch to speedup search performance which includes the page cache, request cache, and query cache \cite{elasticsearch-caching}. The page cache is typically managed by the OS and lacks the semantic understanding of the data being cached. The request cache stores full responses to complex queries (such as aggregations) that are time-consuming to be run again, while the query cache uses heuristics to identify portions of the query responses that may be useful for other queries and stores them selectively. Neither Elasticsearch nor OS provide any APIs to evict select data items, even the deleted ones; so, the only option to remove any item is to clear the entire cache.

\vspace{1mm}
The next two components, transaction log and snapshots, bring fault tolerance to Elasticsearch. Translog is a record of all modifications to Elasticsearch (such as inserts, updates, and deletes) that have been accepted but not yet been committed to the Lucene index on the disk. The idea is to flush changes to the Lucene index in batches, so the overhead is amortized. However, to make this scheme crash consistent (i.e., retain modifications after a process crash or hardware failure), Translog itself has to be continually saved to the disk. Issuing a delete request on a data item that is present in Translog will not remove it from Translog; instead, it will continue to be there until Translog becomes large enough to be flushed (controlled by the parameter \texttt{flush\_threshold\_size}). On the other hand, snapshot is a full backup of an Elasticsearch instance that can be used to transfer Elasticsearch between servers and to restore after a hardware failure. These exist as immutable files on the disk. Unfortunately, there are no APIs to remove select data items from a snapshot. The only way to achieve this would be to restore a snapshot, perform the deletions, and then create a new snapshot in its place.

\vspace{1mm}
The fifth element is the logging mechanism. While Elasticsearch generates several types of logs during its operation, two of them can contain actual data: Elasticsearch.log when set to the level \texttt{trace} and slowlog. By default, these are stored on the disk and are automatically compressed with passage of time. Elasticsearch neither provides an API to search logs that contain a given data item nor offer mechanisms to selectively delete items. However, since log files are in human readable ASCII, standard file I/O operations could be used to accomplish these tasks.

\vspace{1mm}
The final component is the replica shard. Elasticsearch employs a primary-backup model to improve availability. All indexing operations of Elasticsearch are directed to the primary, which is then responsible for pushing any change in its state to all the replicas. On the other hand, read requests can be sent to either primary or one of the replicas, depending on the system load. Thus, replica nodes will have their own data structures, caches, translog, and event logs as discussed in Table \ref{tbl-cleansing-delete}. However, Elasticsearch does not provide any APIs directly manipulate these on the replicas, instead requiring all changes to be driven by the primary (in order to keep them in sync). It must be noted that the delay in propagating the changes from primary to replicas is negligible since these are handled synchronously during API calls.

\vspace{1mm}
\vheading{Implementation.}
Next, we share our effort at introducing cleansing delete for Elasticsearch. We choose to implement this external to Elasticsearch (i.e., using its existing APIs, external scripts and OS support) as opposed to redesigning the internals of Elasticsearch. Since our goal is to demonstrate the feasibility of cleansing delete and estimate the resulting overhead, this approach is sufficient (though we acknowledge that redesigning Elasticsearch would lead to a more optimal solution). Here are our five steps of cleansing: (i) for data structures, we issue \texttt{forcemerge} with \texttt{expunge-deletes} option set, (ii) for cache system, we invoke the OS command \texttt{drop\_caches} and Elasticsearch's index-level clear cache API, (iii) for Translog, we issue Elasticsearch's index-level flush API, (iv) for snapshot, we restore it on a new server, issue delete requests on the select data, flush and forcemerge the index, and then create a new snapshot to replace the old one, (v) finally, for event logs, we iterate through all the files in the log directories and edit out the select data. We implement all these routines in $\sim$ 450 lines of Python code.

\vspace{1mm}
\vheading{Benchmarking.}
We use Chameleon Cloud \cite{Chameleon} for benchmarking. Elasticsearch is run on dedicated Dell PowerEdge R6525 server with 64-core AMD EPYC 2.45GHz processor, 256MB cache, 256GB memory, 480GB SSD storage, and Gigabit Ethernet. We use Elasticsearch version 8.10 and run it in a single replica configuration. We benchmark Elasticsearch's performance using Rally version 2.10 \cite{rally}. In particular, we use the StackOverflow track, which consists of 36 million documents, each of which is a question and answers post from StackOverflow, adding up to a total repository size of 33GB. For comparison, the English Wikipedia has $\sim$62M pages and a size of 58GB. Table \ref{tbl-rally} lists six challenges (or workloads) representing a broad mix of query complexity and the throughput they achieve on our setup. We configure a benchmark run as 200 iterations of warm up followed by 500 iterations of challenge queries.

%--------------------------------------------------------------------
% Elasticsearch components with data 
%--------------------------------------------------------------------
\begin{table}[t]
\caption{\bf Elassticsearch performance (average throughput measured in op/s) against the StackOverflow challenges}
\label{tbl-rally}
{ \renewcommand{\arraystretch}{1.1}
  \begin{tabular}{@{}lll@{}}
    \toprule[1.2pt]
    {Challenge} & {Workload description} & {Throughput} \\
    \midrule
    \multirow{2}{*}{high-low} & boolean AND of high occurrence & \multirow{2}{*}{164.4} \\
     & and low occurrence queries & \\
    prefix & search based on prefix within a field & 73.2 \\
    range & search based on a date range & 55.9 \\
    fuzzy & search based on approx. string match & 23.1 \\
    wildcard & search based on wildcard patterns & 15.0 \\
    \multirow{2}{*}{agg-wc-fz} & wildcard based fuzzy search followed & \multirow{2}{*}{6.7}\\
     & by aggregation & \\
    \bottomrule[1.2pt]
  \end{tabular}}
\vspace{-3mm}
\end{table}
%--------------------------------------------------------------------

\vspace{1mm}
The key goal of this benchmarking is to determine if and how significant is the impact of cleansing deletion on Elasticsearch. To measure the runtime impact of cleansing deletion, we issue two types of RTBF requests while the Rally benchmarking is going on: first, deleting a single document and second, deleting 1K documents. The former represents the practice of deleting data as soon as RTBF is approved while the latter represents batching of RTBF approvals so all data can be deleted at once. Under both these setup, Forcemerge and flush operations complete in less than a millisecond, and we do not see any drop in Elasticsearch throughput\footnote{We must note that the latency of Forcemerge is proportional to the size of deleted data. For instance, if we delete 18M documents (i.e., 50\% of data), then Forcemege takes 23 minutes to complete in our setup. We don't anticipate deletions happening at that scale frequently.}. Next, clearing the cache takes $\sim$10 milliseconds, but it instantaneously reduces the Elasticsearch throughput to 50-60\% of the original under all workloads. The performance soon reverts back to the original levels once the cache becomes hot again (which takes 6 to 8 seconds depending on the challenge). Restoring and creating a snapshot on the full StackOverflow workload takes $\sim$12 minutes each in our setup, while event log culling runs in less than a second. In summary, if we issue a cleansing deletion on a small amount of data (say, not exceeding 1\% of the total dataset), then in our setup, it (i) temporarily reduces the throughput for $\sim$6 seconds, and (ii) takes about 24 minutes to complete. 

\vspace{1mm}
\vheading{Key Takeaway.}
Based on the ease of implementation and a minimal impact on performance, we recommend that cleansing delete be integrated in Elasticsearch, especially if the RTBF workflows could be batch scheduled at times of lower system load, say nightly or weekly.

%\section{Recommendations}
%\label{sec:recommendations}
%Introducing RTBF capability into an application does not mean that it is GDPR compliant; it is just the first step.

%\vheading{Method over madness.}
%TBA.

%\vheading{Prioritizing certain tasks.}
%Not all RTBF tasks are equally visible (reword). Lastly, in the history of GDPR, there is not a single fine, reprimand, or criticism for honoring an RTBF request that should have been rejected. It is quite logical since people tend to complain if their requests are not honored (reword). Please note that we are not recommending a honor-all-RTBF-requests policy but advocating for a calibrated approach to policy management.

%\vheading{RTBF on the rise.}
%Trend over last five years. RTBF and AI/ML systems.

\section{Conclusion}
\label{sec:conclusion}
Ever since its introduction, RTBF has triggered vigorous debates in our society---from being hailed it as a counterbalance to the aggressive data practices of the 21st century to being criticized as a means to rewrite history, weaken the freedom of expression, and enable censorship. This work is an attempt to bring clarity about implementing RTBF from a computing and data management perspective. We believe that our principled approach to understanding the law and enforcement of RTBF, and then translating them to implementable actions in computing systems, will bridge the disconnect between legal and computing domains. 

\vspace{1mm}
\vheading{Acknowledgment.}
Any opinions, findings, or recommendations expressed herein are those of the authors; these should neither be interpreted as legal advice nor as reflective of the views of their host institution.

\bibliographystyle{ACM-Reference-Format}
\bibliography{paper}

\end{document}